\documentclass[aps,prb,showpacs,groupedaddress]{revtex4}
\usepackage{amsmath,graphicx}
\usepackage{epsfig}
\usepackage{bm}
\usepackage[next]{inputenc}

\begin{document}
\title{Thermal conductivity of anisotropic spin - 1/2 two leg ladder:\\Green's function approach}
\author{H. Rezania}
\email[]{rezania.hamed@gmail.com}
\affiliation{Physics Department, Razi Univerity, Kermanshah, Iran}
\author{A. Langari}
%\email[]{langari@sharif.edu}
%\homepage[]{http://spin.cscm.ir}
\affiliation{Department of Physics, Sharif University of Technology, Tehran
11155-9161, Iran, and \\
Center of excellence in Complex Systems and Condensed Matter
(CSCM), Sharif
University of Technology, Tehran 1458889694, Iran
}
\author{Paul H. M. van Loosdrecht}
\affiliation{Zernike Institute for Advanced Materials, University of Groningen, 
Nijenborgh 4, 9747 AG Groningen, The Netherlands, and \\
II. Physikalisches Institut, 
University of Cologne, 50937 K\"{o}ln, Germany}
\author{X. Zotos}
\affiliation{Department of Physics and Institute of Theoretical and Computational
Physics, University of Crete, Greece}

\begin{abstract}
We study the thermal transport of a spin-1/2 two leg antiferromagnetic ladder in the direction of legs. 
The possible effect of spin-orbit coupling and crystalline electric field are investigated in terms of 
anisotropies in the Heisenberg interactions on both leg and rung couplings. The original spin 
ladder is mapped to a bosonic model via a bond-operator transformation where an infinite hard-core
repulsion is imposed to constrain one boson occupation per site. The Green's function approach
is applied to obtain the energy spectrum of quasi-particle excitations responsible
for thermal transport.  The thermal conductivity is found to be monotonically decreasing with temperature
due to increased scattering among triplet excitations at higher temperatures. A tiny 
dependence of thermal transport on the anisotropy in the leg direction at low temperatures is observed 
in contrast to the strong one on the anisotropy along the rung direction, due to the direct  
effect of the triplet density. Our results reach asymptotically the ballistic regime
of the spin - 1/2 Heisenberg chain and compare favorably well with exact diagonalization data.

\end{abstract}

\date{\today}
\maketitle
\section{Introduction}
The spin liquid phase \cite{anderson1,anderson2} has been the focus of numerous theoretical as well as experimental 
studies as it appears in 
both gapless and gapped quasi- one dimensional quantum magnets. For instance, the two leg spin - 1/2 
ladder compounds with standard geometry and antiferromagnetic exchange couplings exhibits a spin liquid state with 
a Haldane type energy gap\cite{dagotto}.
This phase is characterized by exotic magnetic excitations, as the spinons which are topological excitations in the 
spin - 1/2 Heisenberg chain, or the S=1 excitations, commonly called "triplons", in the gapped ladders.
One of the most fascinating manifestations of the magnetic excitations is the observation of a magnetic mode 
of heat transport\cite{hess,sol} in quasi-one dimensional magnets as the Sr$_2$CuO$_3$, SrCuO$_2$ 
chain or the Sr$_{14}$Cu$_{24}$O$_{41}$ ladder compounds\cite{hessrev}.
These are electrically insulating compounds (ceramics) with a large, highly anisotropic heat conductivity 
that is attributed to the propagation of magnetic excitations along the chains (ladders) and of magnitude 
dictated by the magnetic exchange constant that is often of the order of magnitude of the Fermi energy in 
metallic systems.

From the theoretical point of view, a ballistic magnetic heat transport is predicted in exactly solvable 
spin Hamiltonians like the antiferromagnetic spin - 1/2 Heisenberg chain.
\cite{zotos,meisner1,meisner2,orignac,saito}. Very high purity Sr$_2$CuO$_3$, SrCuO$_2$ compounds, which 
are very good realizations of this model as the interchain coupling is very weak, have confirmed this expectation. 
However, due to the presence of spin-phonon scattering, the heat conductivity becomes finite, monotonically 
decreasing with temperature above about 50K \cite{hlubek}. 
The thermal conductivity of spin - 1/2 Heisenberg and XY chains coupled to optical and acoustic phonons 
has been studied by numerical simulations and reveals similar behavior for the two models, with an enhancement
in the Heisenberg model due to larger energy current correlations \cite{louis}. Furthermore, 
bosonization\cite{shimshoni} and a Boltzmann semi-phenomenological approach\cite{chernyshev} yield 
qualitatively similar behavior.
The effect of next-nearest neighbor (n.n.n) interaction and interchain coupling on the 
thermal transport of the anisotropic Heisenberg model has also been investigated by exact diagonalization (ED) on finite
clusters \cite{jung}. This study shows that while the thermal transport of the integrable 
one-dimensional s - 1/2 Heisenberg 
model is infinite at all finite temperatures, both interchain and n.n.n. interactions which 
break the integrability of the model reduce the thermal transport, with the reduction being more
pronounced for the interchain coupling. 

For the spin ladder systems the situation is more complex. Here, the ballistic magnetic thermal conduction is 
limited both by the magnon-magnon scattering as well as the magnon-phonon one (as the elementary spin 
excitations, triplons, carry spin S=1 ), even though the coupling to acoustic phonons is weak \cite{montagnese}. 
What is really surprising and worth understanding 
is that, despite the scattering mechanisms, the thermal conductivity of the ladder compounds is as high as that 
of the spin chain ones\cite{hess1}.
An early numerical study of the magnetic thermal transport in the two leg spin - 1/2 
ladder, neglecting phonons, indicated that the interchain coupling results in diffusive thermal transport,
at least in the high temperature limit\cite{zotos2004}. Diffusive transport is a result of 
the interchain interactions which break the integrability of the decoupled Heisenberg chains.
For undoped ladder compounds, it was found experimentally that the magnetic 
contribution is very large compared to the phononic one. The temperature dependence of the thermal conductivity 
of Sr$_{14}$Cu$_{24}$O$_{41}$ measured along the ladder direction presents two peaks with the higher temperature peak associated 
with the magnetic transport mode. This is in strong contrast to the two orders of magnitude smaller conductivity 
along the rung direction which only shows the low temperature phonon related peak\cite{hess1}. 
The thermal conductivity of spin ladders systems, including phonons and impurities, has been approached 
theoretically mostly by low energy effective models\cite{orignac,boulat} and numerical simulations\cite{znidaric}.

The goal of this work is to sort out the effect of triplon-triplon scattering in limiting the thermal 
conduction along the leg direction. In particular, we study the interchain and anisotropy dependence 
of the thermal conductivity as a function of temperature using the bond operator 
formalism \cite{sachdev1,sachdev2} where the spin model is mapped to a bosonic one with hard core triplon 
repulsion.
The anisotropies account for the eventual effects of spin orbit coupling and the crystalline electric field. 
Although most ladder compounds are described with isotropic exchange interactions, 
the anisotropy plays an important role in some others like (C$_5$H$_12$N)$_2$CuBr$_4$ \cite{Cizmar2010} and
CaCu$_2$O$_3$ \cite{Kiryukhin2001}.
We have implemented Green's function approach to 
calculate the thermal conductivity, i.e. the time ordered energy current correlation.
Although the calculations are tedious and complex we have tried to elaborate the main steps
in different sections and an appendix. In the last section we discuss and analyze our results 
to show how interchain interactions and anisotropies affect the thermal transport. Moreover, a comparison to
exact diagonalization results \cite{zotos2004} is presented. 

%%%%%%%%%%%%%%%%%%%%%%%%%%%%%%%%%%%%%%%%%%%%%%%%%%%%%%%%%%%%%%%%%%%%%%%%%%%

\section{Anisotropic spin Hamiltonian and its bosonic representation}

The anisotropic Heisenberg Hamiltonian describing the two leg spin - 1/2 ladder is given by,
\begin{eqnarray}
H&=&J_{\bot}\sum_{i}(S^{x}_{i}\tau^{x}_{i}+S^{y}_{i}\tau^{y}_{i}+\Delta S^{z}_{i}\tau^{z}_{i})
+J\sum_{i}(\tau^{x}_{i}\tau^{x}_{i+1}+\tau^{y}_{i}\tau^{y}_{i+1}+\delta \tau^{z}_{i}\tau^{z}_{i+1})\nonumber\\
&+&J\sum_{i}(S^{x}_{i}S^{x}_{i+1}+S^{y}_{i}S^{y}_{i+1}+\delta S^{z}_{i}S^{z}_{i+1}),
\label{e1}
\end{eqnarray}
where $S$ and $\tau$ are the spin - 1/2 operators on the respective legs at position $i$. 
$J$ and $J_{\bot}$ correspond to the exchange coupling between nearest neighbor spins 
along legs and rungs, respectively. $\Delta$ and $\delta$ are the anisotropy parameters, where
$\Delta (\delta)$ denotes the strength of the anisotropy on the rungs (legs).

The bond operator formalism \cite{sachdev1,sachdev2} is defined by the following transformations
\begin{eqnarray}
S_{i,\alpha}=\frac{1}{2}(s_i^{\dagger} t_{i,\alpha}+t^{\dag}_{i,\alpha} s_i-i\epsilon_{\alpha\beta\gamma}t_{i,\beta}t_{i,\gamma}),\nonumber\\
\tau_{i,\alpha}=\frac{1}{2}(-s_i^{\dagger} t_{i,\alpha}-t^{\dag}_{i,\alpha} s_i-i\epsilon_{\alpha\beta\gamma}t_{i,\beta}t_{i,\gamma}),
\label{e1.5}
\end{eqnarray}
where any of the spin operators on the $i$-th rung (as a bond) is expressed in terms of the singlet ($s_i$) or 
three flavor triplet ($t_{i,\alpha}$) operators. The singlet and triplet bond operators satisfy bosonic
commutations relations.
As far as the ratio  $J_{\perp}/J$ is nonzero there is a finite gap between the triplet and singlet states.
Thus, the population of singlet bosons is expected to be much higher than the triplet ones, which justifies to 
consider a condensation of singlets. In the presence of a finite gap, 
we neglect the quantum fluctuation of singlet density and
replace the corresponding operators with its mean value, namely $\langle s\rangle=\sqrt{n_{s}}\approx1$.
Using the bond operator transformation the Hamiltonian can be written in terms of 
a bilinear term and a quartic one. The bilinear part is composed of the local terms
and the intersite terms. The local term, which includes on-site interaction, is given by 
\begin{eqnarray}
H_{local}=J_{\bot}\sum_{i,\alpha=x,y}[(\frac{1+\Delta}{2})t_{i,\alpha}^{\dag}t_{i,\alpha}+t_{i,z}^{\dag}t_{i,z}].
\label{a3}
\end{eqnarray}
The intersite part of bilinear term comes from
the interaction between the nearest neighbor spins    
\begin{eqnarray}
H_{bil}^{(2)}=\frac{J}{2}\sum_{\langle i,j\rangle}\sum_{\alpha=x,y}[t_{i,\alpha}(t_{j,\alpha}+t^{\dag}_{j,\alpha})+h.c.]+\sum_{\langle i,j\rangle}\frac{J\delta}{2}[t_{i,z}(t_{j,z}+t^{\dag}_{j,z})+h.c.].
\label{a4}
\end{eqnarray}
There exists another part in the Hamiltonian, composed of quartic terms in the bosonic triplet operators. 
In the low density limit of the bosonic gas, we can neglect the effect of this term on the 
excitation spectrum of the model which is our case here.
To preserve the spin commutation relations we impose a
hard core constraint on the bosonic gas which can be enforced 
by an infinite on-site interaction between bosons
\begin{eqnarray}
H_{U}=U\sum_{\alpha=x,y,z}t^{\dag}_{i,\alpha}t^{\dag}_{i,\beta}t_{i,\beta}t_{i,\alpha},
\;\;\;U\longrightarrow\infty.
\label{a8}
\end{eqnarray}
The infinite strength of interaction between triplet bosons restricts the occupation of each bond with 
only one boson. The implementation of hard-core repulsion which leads to corrections on the interacting triplet
excitations will be discussed in Sec. \ref{hard-core-repulsion}.
In terms of the Fourier space representation of the triplet operators, the bilinear Hamiltonian is given by, 
$\mathcal{H}_{bil}= H_{local}+H_{bil}^{(2)}$
\begin{eqnarray}
\mathcal{H}_{bil}=\sum_{ k,\alpha=x,y,z}A_{k,\alpha}t_{ k,\alpha}^{\dag}t_{ k,\alpha}
+\sum_{k,\alpha=x,y,z}\frac{B_{k,\alpha}}{2}(t_{k,\alpha}^{\dag} t_{ -k,\alpha}^{\dag}+h.c.),
\label{e6}
\end{eqnarray}
in which the coefficients $A, B$ are
\begin{eqnarray}
A_{k,x}=A_{k, y}&=&J_{\bot}(\frac{1+\Delta}{2})+J \cos(k_{x}),\;\;\;
A_{k,z}=J_{\bot}+\delta J \cos(k_{x}), \;\;\;\nonumber\\
B_{k}&=&-J \cos(k_{x}),\;\;\;B_{k,z}=J\delta \cos(k_{x}).
\label{a9}
\end{eqnarray}
The wave vectors $k_{x}$ are considered in the first Brillouin zone of the ladder ($-\pi<k_{x}<\pi$). 
The effect of hard core repulsion ($U \rightarrow \infty$) 
of the interacting Hamiltonian, Eq.(\ref{a8}), is dominant 
over the remaining quartic terms (which have not been presented here). 
Thus, it is sufficient to take into account the effect of 
hard core repulsion on the triplon spectrum and neglect the remaining quartic terms. 
The interacting part of Hamiltonian in terms of Fourier transformation of bosonic 
operators is given by
\begin{eqnarray}
 \mathcal{H}_{U}=U\sum_{k,k',q}\sum_{\alpha=x,y,z}t^{\dag}_{
k+q,\alpha}t^{\dag}_{k'-q,\beta}t_{k',\beta}t_{k,\alpha} \;.
\label{e4}
\end{eqnarray}
Using a unitary Bogoliubov transformation 
$t_{ k,\alpha}=u_{k,\alpha}\tilde{t}_{k,\alpha}-v_{k,\alpha}\tilde{t}_{-k,\alpha}^{\dag}$, 
the bilinear Hamiltonian is simply diagonalized 
\begin{eqnarray}
\mathcal{H}_{bil}=\sum_{k,\alpha}\omega_{\alpha}(k)\tilde{t}^{\dag}_{k,\alpha}\tilde{t}_{k,\alpha},
\label{a10}
\end{eqnarray}
where $\omega_{k,\alpha}=\sqrt{A_{k,\alpha}^{2}-B_{k,\alpha}^{2}}$is the quasi-particle excitation 
spectrum and 
$u^{2}_{k,\alpha} (v^{2}_{k,\alpha})=(-)\frac{1}{2}+\frac{A_{k,\alpha}}{2(\sqrt{A_{k,\alpha}^{2}-B_{k,\alpha}^{2}})}$ 
denote the Bogoliubov coefficients.

%%%%%%%%%%%%%%%%%%%%%%%%%%%%%%%%%%%%%%%%%%%%%%%%%%%%%%%%%%%%%%%%%%%%%%%%%%%

\section{Green's function approach for the Bosonic gas}

The non-interacting normal Green's function for the Hamiltonian of Eq.(\ref{e6}) 
is $g_{n,\alpha}(k,\tau)=-\langle T(t_{k,\alpha}(\tau)t^{\dag}_{k,\alpha}(0))\rangle$ 
and the anomalous Green's function is given by 
$g_{a,\alpha}(k,\tau)=-\langle T(t^{\dag}_{k,\alpha}(\tau)t^{\dag}_{-k,\alpha}(0))\rangle$. 
Fourier transformation of the normal and anomalous Green's functions are written in the following form
\begin{eqnarray}
g_{n,\alpha}(k,i\omega_{n})&=&\frac{u_{k,\alpha}^{2}}{i\omega_{n}-\omega_{k,\alpha}}-\frac{v_{k,\alpha}^{2}}{i\omega_{n}+\omega_{k,\alpha}},\nonumber\\
g_{a,\alpha}(k,i\omega_{n})&=&\frac{u_{k,\alpha}v_{k,\alpha}}{i\omega_{n}-\omega_{k,\alpha}}-\frac{u_{k,\alpha}v_{k,\alpha}}{i\omega_{n}+\omega_{k,\alpha}},
\label{e10.4}
\end{eqnarray}
where $\omega_{n}=\frac{2n\pi}{\beta}$ denotes the bosonic Matsubara frequency.
The perturbative expansion of the interacting Green's function matrix
in the Matsubara notation \cite{mahan} (for each polarization component of the triplons) is
\begin{equation}
\overline{g}(k,i\omega_{n})=\overline{g^{0}}(k,i\omega_{n})
(1-\overline{g^{0}}(k,i\omega_{n})\overline{\Sigma}(k,i\omega_{n}))^{-1}.
\label{a10.5}
\end{equation}
$\overline{g}(k,i\omega_{n})$ and $\overline{\Sigma}(k,i\omega_{n})$ imply
the interacting Green's function and self-energy matrices given by
\begin{eqnarray}
\overline{g}(k,i\omega_{n})= \left(
                              \begin{array}{cc}
                                g_{n}(k,i\omega_{n}) & g_{a}(k,i\omega_{n}) \\
                                g_{a}(k,i\omega_{n}) & g_{n}(-k,-i\omega_{n}) \\ 

  \end{array}
\right),\;\;\;
\overline{\Sigma}(k,i\omega_{n})= \left(
                              \begin{array}{cc}
                                \Sigma_{n}(k,i\omega_{n}) & \Sigma_{a}(k,i\omega_{n}) \\
                                \Sigma_{a}(k,i\omega_{n}) & \Sigma_{n}(-k,-i\omega_{n}) \\

  \end{array}
\right).
\label{e11}
\end{eqnarray}
%Using the Eqs.(\ref{e10.4},\ref{a10.5},\ref{e11}), the interacting normal Green's function can be resulted. 
The single particle retarded Green's function is obtained in the low energy limit of the retarded
self-energy,
\begin{equation}
G_{n,\alpha}^{sp}(k,\omega)= g_{n}(k,i\omega_{n}\longrightarrow\omega+i0^{+})=
\frac{Z_{k,\alpha}U_{k,\alpha}^{2}}{\omega-\Omega_{k,\alpha}+i0^{+}}-\frac{Z_{k,\alpha}V_{k,\alpha}^{2}}{\omega+\Omega_{k,\alpha}+i0^{+}}.
\label{e12}
\end{equation}
The renormalized excitation spectrum and renormalized single particle weight are given by
\begin{eqnarray}
\Omega_{k,\alpha}&=&Z_{k,\alpha}\sqrt{[A_{k,\alpha}+\Re(\Sigma^{Ret}_{n,\alpha}(k,0))]^{2}-[B_{k,\alpha}+\Re(\Sigma^{Ret}_{a,\alpha}(k,0)]^{2})},
\nonumber\\
&&Z_{k,\alpha}^{-1}=1-(\frac{\partial \Re(\Sigma^{Ret}_{n,\alpha})}{\partial \omega})_{\omega=0},\nonumber\\
&&U_{k,\alpha}^{2} (V_{k,\alpha}^{2})=(-)\frac{1}{2}+\frac{Z_{k,\alpha}[A_{k,\alpha}+\Re(\Sigma^{Ret}_{n,\alpha}(k,0))]}{2\Omega_{k,\alpha}}.
\label{e13}
\end{eqnarray}
The renormalized weight constant is the residue of the single particle pole of the
Green's function. In the next step we will take into account the effect of hard core
repulsion on the magnon spectrum.

%%%%%%%%%%%%%%%%%%%%%%%%%%%%%%%%%%%%%%%%%%%%%%%%%%

\section{Effect of hard core repulsion on the triplon excitation \label{hard-core-repulsion}}

The density of the triplons for each polarization ($n_{\alpha}$) component can be easily obtained 
by using the normal Green's functions 
\begin{eqnarray}
n_{\alpha=x,y,z}\equiv\frac{1}{L}\sum_{i}\langle t^{\dag}_{i,\alpha}t_{
i,\alpha}\rangle=\frac{1}{L}\sum_{k}\{[1+2n_{B}(\omega_{k,\alpha})]v^{2}_{k,\alpha}
+n_{B}(\omega_{k,\alpha})\},
\label{e131}
\end{eqnarray}
where $L$ is the number of rungs on the ladder and $n_{B}$ is the Bose-Einstein distribution function.
Since the Hamiltonian $\mathcal{H}_{U}$ in Eq.(\ref{e4}) is short ranged 
and $U$ is large, 
the Brueckner approach (ladder diagram summation) \cite{gorkov,fetter} can be applied
in the low density limit of the bosonic gas and at low temperatures  $T<J_{\bot}, J$.
The interacting normal Green's function is obtained by imposing the hard core boson 
repulsion, $U\rightarrow \infty$. 
Firstly, the scattering amplitude (t-matrix) $\Gamma(k_{1},k_{2};k_{3},k_{4})$ 
of magnons is introduced where $k_{i}\equiv(k,(k_{0}))_{i}$. 
%$\textbf{K}=\textbf{p}_{1}+\textbf{p}_{2}$.
The basic approximation made in the derivation of $\Gamma(P,i\omega_{n})$ is
that we neglect all anomalous scattering vertices, which are
present in the theory due to the existence of anomalous Green's functions.
According to the Feynman rules \cite{fetter}, in momentum space at finite temperature 
and after taking limit $U\longrightarrow\infty$, the scattering amplitude is written by 
(see Fig.1 of Ref.\onlinecite{rezania2008})
\begin{eqnarray}
\Gamma_{\alpha\beta,\alpha\beta}(P,i\omega_{n})=\Big(\frac{1}{\beta2\pi}\sum_{m}\int dQ G^{(0)}_{\alpha\alpha}(Q,iQ_{m})G^{(0)}_{\beta\beta}(P-Q,i\omega_{n}
-iQ_{m})\Big)^{-1}.
\label{a13}
\end{eqnarray}
where,  $p_{1}+p_{2}=p_{3}+p_{4}\equiv(P,i\omega_{n})$.
However, the key observation is that all anomalous contributions
are suppressed by an additional small parameter present in the
theory.
Indeed, both terms of the anomalous scattering matrix are proportional to 
$v_{q}^{2}$ (which is proportional to the density of triplons) and therefore can be neglected. 
We find the solution self-consistently putting 
$G^{0}\longrightarrow G$ in Eq.(\ref{a13}).
By replacing the  non-interacting normal Green's function 
%(Eq.(\ref{e69},\ref{e70}))
in the Bethe-Salpeter equation (Eq.(\ref{a13})), taking the 
limit $U\longrightarrow\infty$ and in addition considering the fluctuation-dissipation theorem 
in which the Matsubara representation of the Green's function 
is related to the spectral function 
($G_{\alpha}(k,i\omega_{n})=\int_{-\infty}^{\infty}\frac{d\omega}{2\pi}\frac{-2Im[G_{\alpha}^{Ret}(k,\omega)]}{i\omega_{n}-\omega}$), we obtain 
the scattering matrix in the following form, 
\begin{eqnarray}
\Gamma_{\alpha\beta,\alpha\beta}(P,i\omega_{n})=-\Big(\frac{1}
{2\pi}
\int dQ &[&u^{2}_{Q,\alpha}u^{2}_{P-Q,\beta}
(\frac{n_{B}(\omega_{Q,\alpha})}{i\omega_{n}-\omega_{Q,\alpha}-\omega_{P-Q,\beta}}
-\frac{n_{B}(-\omega_{P-Q,\beta})}{i\omega_{n}-\omega_{P-Q,\beta}-\omega_{Q,\alpha}})\nonumber\\
&-&u^{2}_{Q,\alpha}v^{2}_{P-Q,\beta}(\frac{n_{B}(\omega_{Q,\alpha})}{i\omega_{n}-\omega_{Q,\alpha}+\omega_{P-Q,\beta}}
-\frac{n_{B}(\omega_{P-Q,\beta})}{i\omega_{n}+\omega_{P-Q,\beta}-\omega_{Q,\alpha}})\nonumber\\
&-&v^{2}_{Q,\alpha}u^{2}_{K-Q,\beta}
(\frac{n_{B}(-\omega_{Q,\alpha})}{i\omega_{n}+\omega_{Q,\alpha}-\omega_{K-Q,\beta}}
-\frac{n_{B}(-\omega_{P-Q,\beta})}{i\omega_{n}-\omega_{P-Q,\beta}+\omega_{Q,\alpha}})\nonumber\\
&+&v^{2}_{Q,\alpha}v^{2}_{P-Q,\beta}
(\frac{n_{B}(-\omega_{Q,\alpha})}{i\omega_{n}+\omega_{Q,\alpha}+\omega_{P-Q,\beta}}
-\frac{n_{B}(\omega_{P-Q,\beta})}{i\omega_{n}+\omega_{P-Q,\beta}+\omega_{Q,\alpha}})]
\Big)^{-1}.
 \label{e178}
\end{eqnarray}
The low density limit of the bosonic gas implies that we can neglect terms including the 
coefficients $v_{\alpha}$.
%we should perform summation over internal Matsubara frequencies for interacting Green's functions. 
%Finally, to replace Matsubara Green's function by retarded Green's function we can write 
%interacting retarded Green's function similar to non-interacting Green's function by renormalization 
%the parameters($u\longrightarrow\sqrt{Z}U,v\longrightarrow\sqrt{Z}V,\omega\longrightarrow\Omega$)    exists between 
%Because of the strong interaction between the triplet bosons we should carry out the expansion in
%Dyson's equation to infinite order. Now ,we calculate summation over Matsubara frequency in the Eq. (\ref{e170}).
%The zero temperature limit, $\beta\longrightarrow\infty$ of Eq.(\ref{e178}) reproduces the scattering 
%amplitude in Ref.\onlinecite{rezania}.
%%%%%%%%%%%%%%%%%%%%%%%%%%%%%%%%%%%%%%%%%%%%%%%%%%%%%%%%%%%%%
%\begin{figure}
%\includegraphics[width=8cm]{fig2.eps}
%\hspace{2pc}%
%\begin{minipage}[b]{5.5cm}
%\caption{\label{fig2}The anomalous (top) and normal (bottom) self-energy diagrams. }
%\end{minipage}
%\end{figure}
%%%%%%%%%%%%%%%%%%%%%%%%%%%%%%%%%%%%%%%%%%%%%%%%%%%%%%%%%%%
According to Fig.2 of Ref.\onlinecite{rezania2008}, the normal self-energy is obtained by using the vertex-function 
obtained in Eq.(\ref{e178})
\begin{eqnarray}
\Sigma^{U}_{\alpha\alpha}(k,i\omega_{n})&=&-\sum_{p_{m},\gamma}\int^{\infty}_{-\infty}\frac{dp}{2\pi\beta}
\Gamma_{\alpha\gamma,\alpha\gamma}(p+k,i\omega_{n}+ip_{m})G_{\gamma\gamma}(p,ip_{m})\nonumber\\&-&\sum_{p_{m}}\int^{\infty}_{-\infty}\frac{dp}{2\pi\beta}
\Gamma_{\alpha\alpha,\alpha\alpha}(p+k,i\omega_{n}+ip_{m})G_{\alpha\alpha}(p,ip_{m}). \hspace{3mm}
\label{e190}
\end{eqnarray}
After performing the integration
on the internal Matsubara frequency ($p_{m}$), the normal self-energy is obtained in the following form 
\begin{eqnarray}
\Sigma^{U}_{xx}(k,i\omega_{n})&&=\frac{3}{2\pi}\int dp\Big(u_{p,x}^{2}n_{B}(\omega_{p,x})\Gamma_{xx,xx}(p+k,\omega_{p,x}+i\omega_{n})-
v_{p,x}^{2}n_{B}(-\omega_{p,x})\Gamma_{xx,xx}(p+k,-\omega_{p,x}+i\omega_{n})\Big)\nonumber\\
+&&\frac{1}{2\pi}\int dp\Big(u_{p,z}^{2}n_{B}(\omega_{p,z})\Gamma_{xz,xz}(p+k,\omega_{p,z}+i\omega_{n})-v_{p,z}^{2}n_{B}
(-\omega_{p,z})\Gamma_{xz,xz}(p+k,-\omega_{p,z}+i\omega_{n})\Big).
 \label{e211}
\end{eqnarray}
The other components of the self-energy are found in a similar way. In addition to the normal self-energy 
presented in Eq.(\ref{e211}), there are anomalous self-energy diagrams which are formally at most linear in 
the density of the bosonic gas.
 In the dilute gas approximation, the contributions of 
such terms are numerically smaller than  Eq.~(\ref{e211}).

%%%%%%%%%%%%%%%%%%%%%%%%%%%%%%%%%%%%%%%%%%%%%%%%%%%%%%%%%%%%%%%%%%%%%%%%%%%%%%%

\section{Energy current and thermal conductivity}

The thermal conductivity is obtained as the response of the energy current (${\bf{J}}_{E}$) 
to a temperature gradient. Imposing the continuity equation for the energy density, 
$\frac{\partial}{\partial t}H+\nabla \cdot \textbf{J}_{E}=0$, 
the explicit form of the energy current can be calculated. The Hamiltonian can be considered
as a sum of local Hamiltonians $H=\sum_{m}h_{m}$ in which the local terms ($h_m$) are
\begin{eqnarray}
h_{m}&=&J(S^{x}_{m}S^{x}_{m+1}+S^{y}_{m}S^{y}_{m+1}+\delta S^{z}_{m}S^{z}_{m+1}
+\tau^{x}_{m}\tau^{x}_{m+1}+\tau^{y}_{m}\tau^{y}_{m+1}+\delta\tau^{z}_{m}\tau^{z}_{m+1})
\nonumber\\&+&J_{\bot}(S_{m}^{x}\tau_{m}^{x}+S_{m}^{y}\tau_{m}^{y}+\Delta S_{m}^{z}\tau_{m}^{z}).
%-(S_{m}^{z}+\tau_{m}^{z})B_{z},
\label{e214}
\end{eqnarray}
The energy current can be derived formally by defining an operator which is the summation 
over the position vector and the local Hamiltonian
\begin{eqnarray}
\textbf{R}_{E}\equiv\sum_{i}\textbf{R}_{i}h_{i},
\label{e212}
\end{eqnarray}
where $h_{i}$ has been introduced in Eq.(\ref{e214}) and ${\bf R}_i$ denotes the position of 
a rung on the lattice. Using the continuity equation, the energy current operator is reduced to
\begin{eqnarray}
{\bf{J}}_{E}=\frac{\partial}{\partial t}\textbf{R}_{E}
=\sum_{l}\textbf{R}_{l}\frac{\partial}{\partial t}h_{l}=i\sum_{l,m}\textbf{R}_{l}[h_{m},h_{l}].
\label{e213}
\end{eqnarray}
 After some calculations, the component of the energy current along the $x$ direction is given by, 
\begin{eqnarray}
{J}^{x}_{E}=&J^{2}&\sum_{m}\Big(-S_{m}^{x}S_{m+a}^{z}S^{y}_{m+2a}+S_{m}^{y}S^{z}_{m+a}S^{x}_{m+2a}+\delta(S_{m}^{x}S_{m+a}^{y}S_{m+2a}^{z}-S_{m}^{z}S_{m+a}^{y}S_{m+2a}^{x})\nonumber\\&+&\delta(S_{m}^{z}S_{m+a}^{x}S_{m+2a}^{y}-S_{m}^{y}S_{m+a}^{x}S_{m+2a}^{z})+(S\longrightarrow\tau)\Big)+J_{\bot}J\sum_{m}\Big(S_{m}^{y}S_{m+a}^{z}\tau^{x}_{m+a}-S_{m}^{x}S_{m+a}^{z}\tau^{y}_{m+a}\nonumber\\&+&\delta(S_{m}^{z}S_{m+a}^{x}\tau_{m+a}^{y}-S_{m}^{z}S_{m+a}^{y}\tau_{m+a}^{x})+\Delta(S_{m}^{x}S_{m+a}^{y}\tau_{m+a}^{z}-S_{m}^{y}S_{m+a}^{x}\tau_{m+a}^{z})
+( S\longrightarrow \tau, \tau\longrightarrow S)\Big).
\label{e215}
\end{eqnarray}
The above equation can be rewritten in terms of Fourier transformation of spin operators
\begin{eqnarray}
{J}^{x}_{E}=&J^{2}&\frac{1}{L^{2}}\sum_{q,q^{'}}e^{-2iq_{x}-iq^{'}_{x}}\Big(-S_{q}^{x}S_{q^{'}}^{z}S^{y}_{-(q+q^{'})}+S_{q}^{y}S^{z}_{q^{'}}S^{x}_{-(q+q^{'})}+\delta(S_{q}^{x}S_{q^{'}}^{y}S_{-(q+q^{'})}^{z}-S_{q}^{z}S_{q^{'}}^{y}S_{-(q+q^{'})}^{x})\nonumber\\&+&\delta(S_{q}^{z}S_{q^{'}}^{x}S_{-(q+q^{'})}^{y}-S_{q}^{y}S_{q^{'}}^{x}S_{-(q+q^{'})}^{z})+(S\longrightarrow\tau)\Big)+J_{\bot}J\frac{1}{L^{2}}\sum_{q,q^{'}}e^{-iq_{x}}\Big(S_{q}^{y}S_{q^{'}}^{z}\tau^{x}_{-(q+q^{'})}-S_{q}^{x}S_{q^{'}}^{z}\tau^{y}_{-(q+q^{'})}\nonumber\\&+&\delta(S_{q}^{z}S_{q^{'}}^{x}\tau_{-(q+q^{'})}^{y}-S_{q}^{z}S_{q^{'}}^{y}\tau_{-(q+q^{'})}^{x})+\Delta(S_{q}^{x}S_{q^{'}}^{y}\tau_{-(q+q^{'})}^{z}-S_{q}^{y}S_{q^{'}}^{x}\tau_{-(q+q^{'})}^{z})+(S\longrightarrow\tau,\tau\longrightarrow S)\Big),
\label{e216}
\end{eqnarray}
where $L$ is the number of rungs. 
The linear response theory is implemented to obtain the thermal conductivity under the assumption of
a low temperature gradient (as a perturbing field).
The Kubo formula gives the transport coefficient $L_{22}(\omega)$ in terms of a correlation function 
of energy current operators 
\begin{eqnarray}
L^{Ret}_{22}(\omega)=\frac{i}{\beta\omega}\int_{-\infty}^{+\infty}dt e^{i\omega t}\theta(t)\langle[j^{x}_{E}(t),j^{x}_{E}(0)]\rangle=
\frac{1}{\beta\omega}\lim_{i\omega_{n}\longrightarrow\omega+i0^{+}}\int^{\beta}_{0}d\tau e^{i\omega_{n}\tau}\langle T_{\tau}(j^{x}_{E}(\tau)j^{x}_{E}(0))\rangle
\label{e216.5}
\end{eqnarray}
The energy current density is related to the temperature gradient via ${\bf J}_{E}=-K\nabla T$ where $K$ 
is the transport coefficient \cite{mahan,grosso}.
%  Thus the energy flows in the direction opposite to $\nabla T$. 
%The thermal coefficient is related to the transport coefficient by 
%$K=\frac{1}{T^{2}}L_{22}(\omega\longrightarrow 0)$
The thermal conductivity and 
$L_{22}$ are related by\cite{kotliar}
\begin{eqnarray}
K=-\beta^{2}\lim_{\omega\longrightarrow 0}\Im(L^{Ret}_{22}(\omega)).
\label{e217}
\end{eqnarray}
We calculate the correlation function in Eq.(\ref{e216.5}) within an approximation 
by implementing Wick's theorem. 
The correlation functions between current operators can be interpreted as 
multiplication of three dynamical spin susceptibilities in the form
\begin{eqnarray}
\kappa(\tau)\equiv\langle T_{\tau}(j^{x}_{E}(\tau)j^{x}_{E}(0))\rangle&=&\frac{J^{4}}{L^{4}}\sum_{k,k^{'},k_{1},k^{'}_{1}}\sum_{\alpha\beta\alpha^{'}\beta^{'}\gamma\gamma^{'}}\epsilon_{\alpha\beta\gamma}\epsilon_{\alpha^{'}\beta^{'}\gamma^{'}} e^{-2ik_{x}-ik^{'}_{x}-2ik_{1x}-ik^{'}_{1x}}\times\nonumber\\&&\langle T(S^{\alpha}_{k}(\tau)S_{k^{'}}^{\beta}(\tau)S_{-(k+k^{'})}^{\gamma}(\tau)S_{k_{1}}^{\alpha^{'}}(0)S_{k^{'}_{1}}^{\beta^{'}}(0)S_{-(k_{1}+k^{'}_{1})}^{\gamma^{'}}(0))\rangle,
\label{e218}
\end{eqnarray}
where $\epsilon_{\alpha\beta\gamma}$ is Levi-Civita tensor.
Applying Wick's theorem, the expectation values in  Eq.(\ref{e218}) are simplified in 
the following form
\begin{eqnarray}
\kappa(\tau)&=&-\frac{12J^{4}}{L^{2}}\sum_{k,k^{'}}\zeta(k_{x},k_{x}^{'})\chi^{xx}(k_{x},\tau)\chi^{yy}(k^{'}_{x},\tau)\chi^{zz}(k^{'}_{x}+k_{x},\tau),\nonumber\\
\zeta(k_{x},k_{x}^{'})&\equiv&1+\delta e^{-3ik_{x}}+\delta^{2}e^{-3i(k_{x}+k_{x}^{'})}-\delta^{2}e^{i(k_{x}^{'}-k_{x})}-\delta e^{-i(k_{x}+2k_{x}^{'})}-\delta^{2}e^{-2i(k_{x}^{'}+2k_{x})},
\label{e220}
\end{eqnarray}
where $\chi(q,\tau)=-\langle T(S_{\alpha}(q,\tau)S_{\alpha}(-q,0))\rangle
=-\langle T(\tau_{\alpha}(q,\tau)\tau_{\alpha}(-q,0))\rangle$ is the dynamical spin correlation 
function. Our calculations indicate that the correlation 
functions including three spin operators such as 
$\langle T(SS\tau)\rangle$ or $\langle T(\tau \tau S)\rangle$ vanish 
and do not contribute to the thermal conductivity. 
%Fourier transformation of $\kappa$ is given by
%\begin{eqnarray}
%\kappa(i\omega_{n})&=&\int^{\beta}_{0}e^{i\omega_{n}\tau}\kappa(\tau)d\tau,\nonumber\\
%\chi_{\alpha}(k,\tau)&=&\frac{1}{\beta}\sum_{n}\chi_{\alpha}(k,i\omega_{n})e^{-i\omega_{n}\tau},\nonumber\\
%\kappa(i\omega_{n})&=&\frac{-12J^{4}}{L^{2}\beta^{2}}\sum_{k,k^{'},n_{1},n_{2}}\zeta(k_{x},k_{x}^{'})\chi^{xx}(k,i\omega_{n_{1}})\chi^{yy}(k^{'},i\omega_{n_{2}})\chi^{zz}(k^{'}+k,i\omega_{n}-i\omega_{n_{1}}-i\omega_{n_{2}})
%\label{e230}
%\end{eqnarray}
The spin susceptibility ($\chi$) obtained by the bond operator transformation 
(Eq.(\ref{e1.5})) is given by 
\begin{eqnarray}
&&\chi_{zz}(k,i\omega_{n})=-\int^{\beta}_{0}d\tau e^{i\omega_{n}\tau}\langle T(S_{z}(k,\tau)S_{z}(-k,0))\rangle\nonumber\\&=&\frac{1}{4}\int^{\beta}_{0}d\tau e^{i\omega_{n}\tau}\nonumber\\
&&\Big\langle T\Big(t_{-k,z}(\tau)+t^{\dag}_{k,z}(\tau)+\sum_{q}(-it^{\dag}_{k+q,x}(\tau)t_{q,y}(\tau)+it^{\dag}_{k+q,y}(\tau)t_{q,x}(\tau))\Big)\times\nonumber\\&&\Big(t_{k,z}(0)+t^{\dag}_{-k,z}(0)+\sum_{q^{'}}(-it^{\dag}_{k-q^{'},x}(0)t_{q^{'},y}(0)+it^{\dag}_{q^{'}-k,y}(0)t_{q^{'},x}(0))\Big)\Big\rangle.
\label{e231}
\end{eqnarray}
Both one and two particle Green's functions contribute to the spin susceptibility. 
Since the anomalous Greens function is negligible compared to the normal Green's function, 
we only consider bubble diagrams that include normal Green's function. The details of the calculation 
of the spin susceptibility via Green's functions of the triplon gas can be found in the Ref.(\onlinecite{rezania2009}).  
After some calculations the z-component of the susceptibility takes the following form
\begin{eqnarray}
\chi_{zz}(k,i\omega_{n})=\frac{1}{4}\Big( u_{k,z}^{2}(\frac{1}{i\omega_{n}-\omega_{k,z}}-\frac{1}{i\omega_{n}+\omega_{k,z}})-2u_{q,x}^{2}u_{k+q,x}^{2}\frac{n_{B}(\omega_{q,x})-n_{B}(\omega_{k+q,x})}{i\omega_{n}-\omega_{k+q,x}+\omega_{q,x}}\Big).
\label{e235}
\end{eqnarray}
In a similar way, the transverse spin susceptibility can be obtained as
\begin{eqnarray}
\chi_{xx}(k,i\omega_{n})=\frac{1}{4}\Big( u_{k,x}^{2}(\frac{1}{i\omega_{n}-\omega_{k,x}}-\frac{1}{i\omega_{n}+\omega_{k,x}})&-&\sum_{q}u_{q,z}^{2}u_{k+q,x}^{2}\frac{n_{B}(\omega_{q,z})-n_{B}(\omega_{k+q,x})}{i\omega_{n}-\omega_{k+q,x}+\omega_{q,z}}\nonumber\\&-&u_{q,x}^{2}u_{k+q,z}^{2}\frac{n_{B}(\omega_{q,x})-n_{B}(\omega_{k+q,z})}{i\omega_{n}-\omega_{k+q,z}+\omega_{q,x}}\Big).
\label{e236}
\end{eqnarray}
In the low density limit of triplons, the terms being proportional to fourth order
in $u$ give the dominant contributions to the spin susceptibilities in the Eqs.(\ref{e235}, \ref{e236}). 
Furthermore, $u$ is close to 1 since $u^2=1+v^2$ and $v^2$ is proportional to the triplon density.
%After substitution of the Eq.(\ref{e236})into Eq.(\ref{e230}) and summation over internal matsubara 
%frequency, the dynamical correlation function ($\kappa(i\omega_{n})$) can be resulted.
Finally, the static thermal conductivity is related to $\kappa(i\omega_{n})$ as
\begin{eqnarray}
K(T)=-\beta \lim_{\omega\longrightarrow0}\frac{1}{\omega}\Im\Big(\kappa(i\omega_{n}\longrightarrow\omega+i0^{+})\Big).
\label{e237}
\end{eqnarray}
The final expression for the static thermal conductivity is given by,
\begin{eqnarray}
K(T)=\frac{12J^{4}\beta}{L^{2}}\sum_{k, k'}\int_{-\infty}^{\infty}\frac{d\epsilon}{2\pi}
\mathcal{D}(k, k', \epsilon)(-\frac{dn_{B}(\epsilon)}{d\epsilon}).
\label{e238}
\end{eqnarray}
The expression for $\mathcal{D}(k, k',\epsilon)$ 
is quite lengthy and is presented in Appendix (A). 
Finally, the thermal conductivity is calculated from the expression given in Eq.(\ref{e238}).

%%%%%%%%%%%%%%%%%%%%%%%%%%%%%%%%%%%%%%%%%%%%%%%%%%%%%%%%%%%%%%%%%%%%%%%%%%%%%%%%%%%%%%%%%%%%%%

\section{Results and discussions}

We have obtained the thermal conductivity of the two leg spin-1/2 antiferromagnetic ladder
along the leg direction in presence of both rung ($\Delta$) and leg ($\delta$) anisotropies.
We have implemented a bosonic representation for the spin ladder where each rung is represented
by bosonic bond-operators , {\em i.e.} a singlet and three flavor triplets.
To preserve the SU(2) spin algebra, a hard core repulsion constraint is added to the bosonic model 
to avoid double occupation of bosons at each lattice site. 
In the limit $J_{\bot}/J\longrightarrow\infty$, 
the spin ladder has a spin liquid ground state which is a direct product of 
singlet states with a finite energy gap to the lowest 
excited state. The energy gap is robust and remains finite even for small values of $J_{\bot}/J$
that defines the energy scale of the quasi-particles called triplons. 
We have obtained the single particle excitations of the bosonic model by means of a Green's 
function approach which gives the thermal conductivity by calculating the
energy current correlation function.
The spin excitations of a 
ladder are calculated within a self-consistent solution of Eqs.(\ref{e211}, \ref{e178}, \ref{e13}), i.e.
by the substitutions $u_{k,\alpha}\longrightarrow \sqrt{Z_{k,\alpha}}U_{k,\alpha}, \;
v_{k,\alpha}\longrightarrow \sqrt{Z_{k,\alpha}}V_{k,\alpha}, \;
%\omega_{-(+)}(k)\longrightarrow\Omega_{-(+)}(k), \;
\omega_{k,\alpha}\longrightarrow\Omega_{k,\alpha}$ into the corresponding equations.
We start with a guess for $Z_{k,\alpha}$, $\Sigma_{\alpha}(k,0)$, 
and find the updated excitations and the renormalized Bogoliubov coefficients using Eq.(\ref{e13}). 
Within a self-consistent iteration we obtain the self-energies and the renormalized 
quasi-particle excitations which leads to the final value of thermal conductivity 
expressed in Eqs.(\ref{e238}, \ref{e239}).

\begin{figure}[ht!]
\begin{center}
\includegraphics[width=10cm]{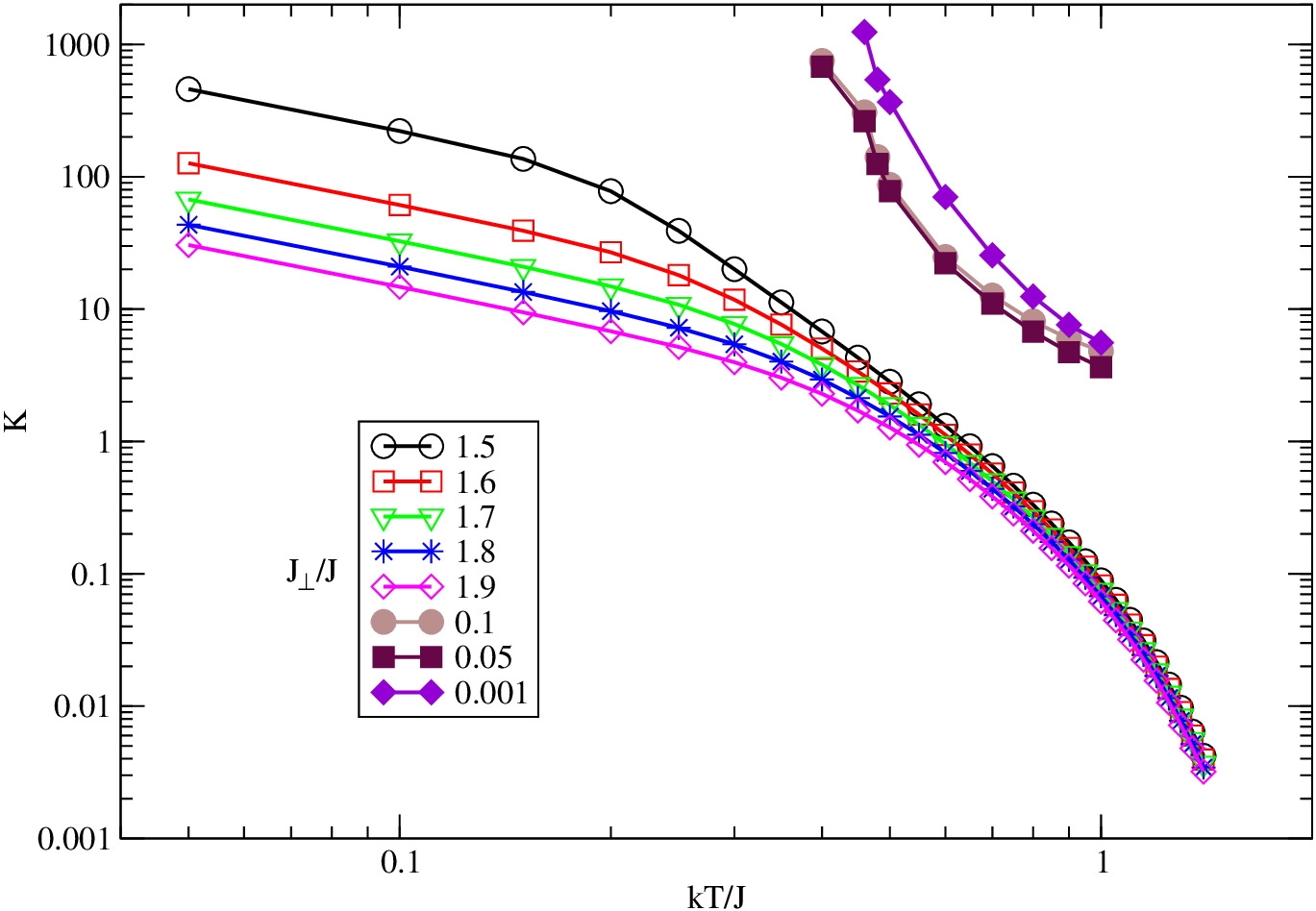}
\end{center}
%\hspace{2pc}%
%\begin{minip-age}[b]{5.5cm}
\caption{\label{fig1} 
Thermal conductivity along leg direction versus $kT/J$ for different values of $J_{\bot}/J$ 
in the isotropic spin ladder ($\Delta=\delta=1$). 
Open symbols correspond to the strong coupling limit $J_{\bot}/J >1$ and filled symbols correspond
to weak coupling $J_{\bot}/J <1$ which asymptotically reaches the ballistic chain limit by
decreasing $J_{\bot}/J$ to $0.001$. 
A monotonic decrease versus temperature 
is seen for all cases.}
%\end{minipage}
\end{figure}

\begin{figure}[ht!]
\begin{center}
\includegraphics[width=10cm]{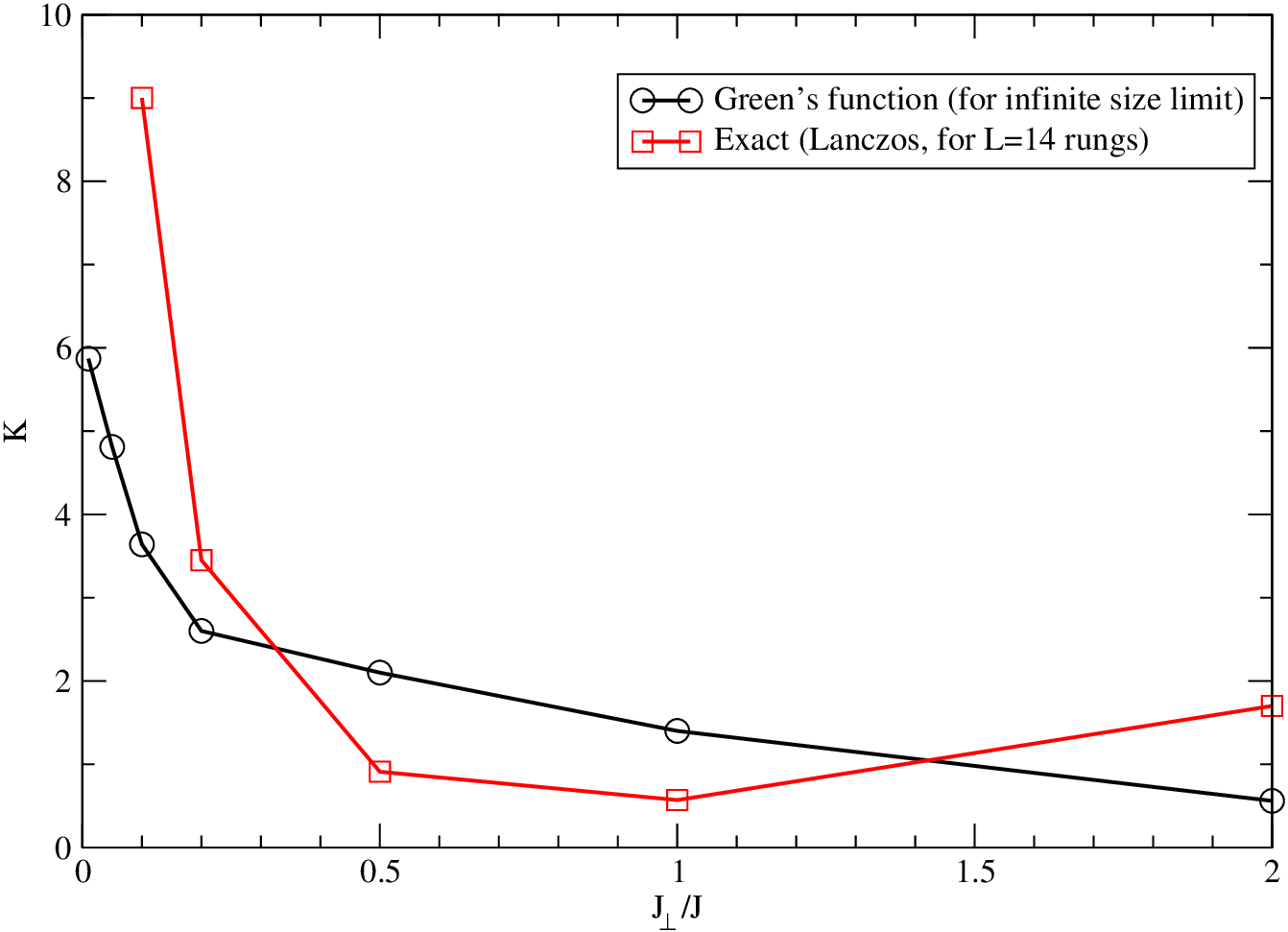}
\end{center}
%\hspace{2pc}%
%\begin{minip-age}[b]{5.5cm}
\caption{\label{fig2} Thermal conductivity as a function of $J_{\bot}/J$ for $kT/J=1$, a comparison between Green's function 
and exact diagonalization \cite{zotos2004} approaches in the isotropic case ($\delta=\Delta=1$). The Green's function result
represent an infinite rung ladder while the exact diagonalization one is done on 14 rungs.}
%\end{minipage}
\end{figure}

In Fig.\ref{fig1} we present the thermal conductivity ($K$) of the isotropic ladder ($\Delta=\delta=1$) 
versus normalized temperature (k$_BT/J$, k$_B$ is Boltzmann constant) for different values 
of rung coupling ($J_{\bot}/J$). Two features are pronounced in this figure. 
The increase of temperature reduces $K$, moreover at 
a fixed temperature the increase of rung coupling leads to a decrease of $K$.
Thermal transport in the two leg ladder is performed via the quasi-particle excitations 
called triplons. Higher temperature causes more scattering of triplons which reduces 
the thermal conductivity. 
A similar result has been reported for dimerized spin chains using the exact 
diagonalization (ED) method \cite{langer}. The decrease of $K$ with temperature is in agreement with 
an experimental study on the two leg ladder at zero hole doping \cite{hess1}.
The increase of rung exchange coupling enhances the
energy gap between the singlet and triplet states on each rung which consequently reduces the number
of triplons that participate in thermal transport and results in lower thermal conductivity.
Conversely, the decrease of rung coupling enhances thermal transport which asymptotically 
diverges as the rung coupling tends to zero as shown in Fig.\ref{fig1} for $J_{\bot}/J=0.001$.
This is in agreement with the ballistic transport of the integrable s - 1/2 Heisenberg chain
expected in the limit $J_{\bot} \rightarrow 0$. This behavior has been observed
in an exact diagonalization study on the two leg ladder \cite{zotos2004} and is compared with
our results in Fig.\ref{fig2}.
The Green's function 
approach data present the behavior of a two leg ladder in the infinite size limit while the ED
ones are obtained from a 14 rung system in the high temperature limit by setting 
k$_B$T/J=1. As the thermal scattering length is expected to be very short in the high temperature limit - 
as evidenced by the absence of finite size dependence\cite{zotos2004} - we expect a qualitative   
agreement and a similar trend in $K$ versus rung coupling as shown in Fig.\ref{fig2}.
The Green's function approach presented here gives accurate results at low temperatures
(k$_B$T/J $\lesssim 0.1$) while the ED results \cite{zotos2004} are justified at high temperatures.
This explains the quantitative difference between the two approaches in Fig.\ref{fig2} while
a qualitative agreement of the trend of results is observed.

We would like to add few comments on the accuracy of our results, which are based on the
two particle scattering matrix ($\Gamma$) that determines the self-energies and consequently
the corrections on the spin excitation spectrum.
$\Gamma$-function which has been presented in Eq.(\ref{e178})
 has been calculated using ladder diagrams
based on the Brueckner approach. 
According to the Brueckner approach, which is justified in the low density limit of bosonic gas,
the normal one particle Green's functions
is the only part of Green's functions which
 constructs the Feynman diagrams of the self-energy and $\Gamma$, which 
means that the anomalous Green's function can be neglected.
%The normal Green's function give dominant contribution to the Feynman diagrams compared to the anomalous one.
This is justified -based on Eq.(\ref{e10.4})- in which one of the terms in the normal Green's function is proportional 
to Bogoliubov coefficient $u^{2}$. While both terms of the anomalous Green's function show a 
dependence on the other Bogoliubov coefficient $v$.
Moreover, Eq.(\ref{e131}) implies that the Bogoliubov coefficient $v$ is negligible in the low density
limit of triplet bosons. Therefore, the factors that increase the density of triplet particles 
reduce the accuracy of our approach.
As a result, one can point out both conditions, the low temperature and high values of $J_{\perp}/J$
yield our approach works more accurately. The former is obvious while the latter case 
increases the gap in the excitation spectrum, which leads to more justified scheme of low triplet density.
As a measure, the density of triplet bosons starts an incremental behavior at 
k$_B$T/J$\sim 0.1$, which proposes to indicate (k$_B$T/J)$\lesssim 0.1$ as the regime in which
our approach gives accurate results. 
Similarly, we would like to consider $(J_{\perp}/J) \gtrsim 0.6$,
where the density of triplets is low enough ($\sum_{\alpha=x,y,z} n_{\alpha} \lesssim 0.1$) 
for a reasonable accuracy.

\begin{figure}[ht!]
\begin{center}
\includegraphics[width=10cm]{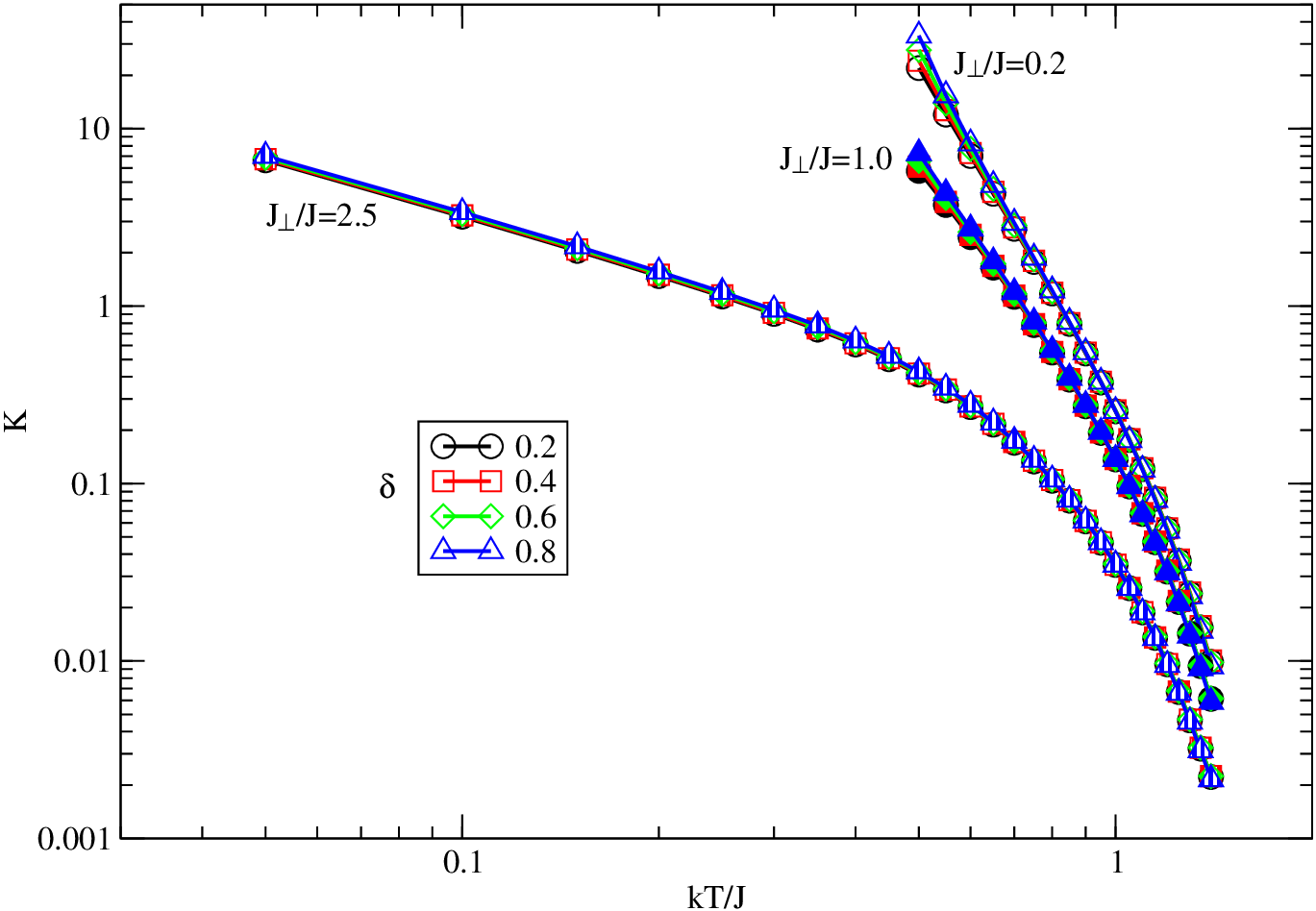}
\end{center}
%\hspace{2pc}%
%\begin{minip-age}[b]{5.5cm}
\caption{\label{fig3} Variation of the thermal conductivity versus normalized temperature ($kT/J$) for 
different anisotropies in the leg direction ($\delta$) and for  $J_{\bot}/J=0.2, 1.0, 2.5$, $\Delta=1$.
Thermal conductivity is increased by decreasing $J_{\bot}/J$.
Open symbols represent $J_{\bot}/J=0.2$, filled symbols correspond to $J_{\bot}/J=1.0$ and the
striped symbols for $J_{\bot}/J=2.5$
The weak dependence on anisotropy becomes observable at low temperatures.}
%\end{minipage}
\end{figure}

\begin{figure}[ht!]
\begin{center}
\includegraphics[width=10cm]{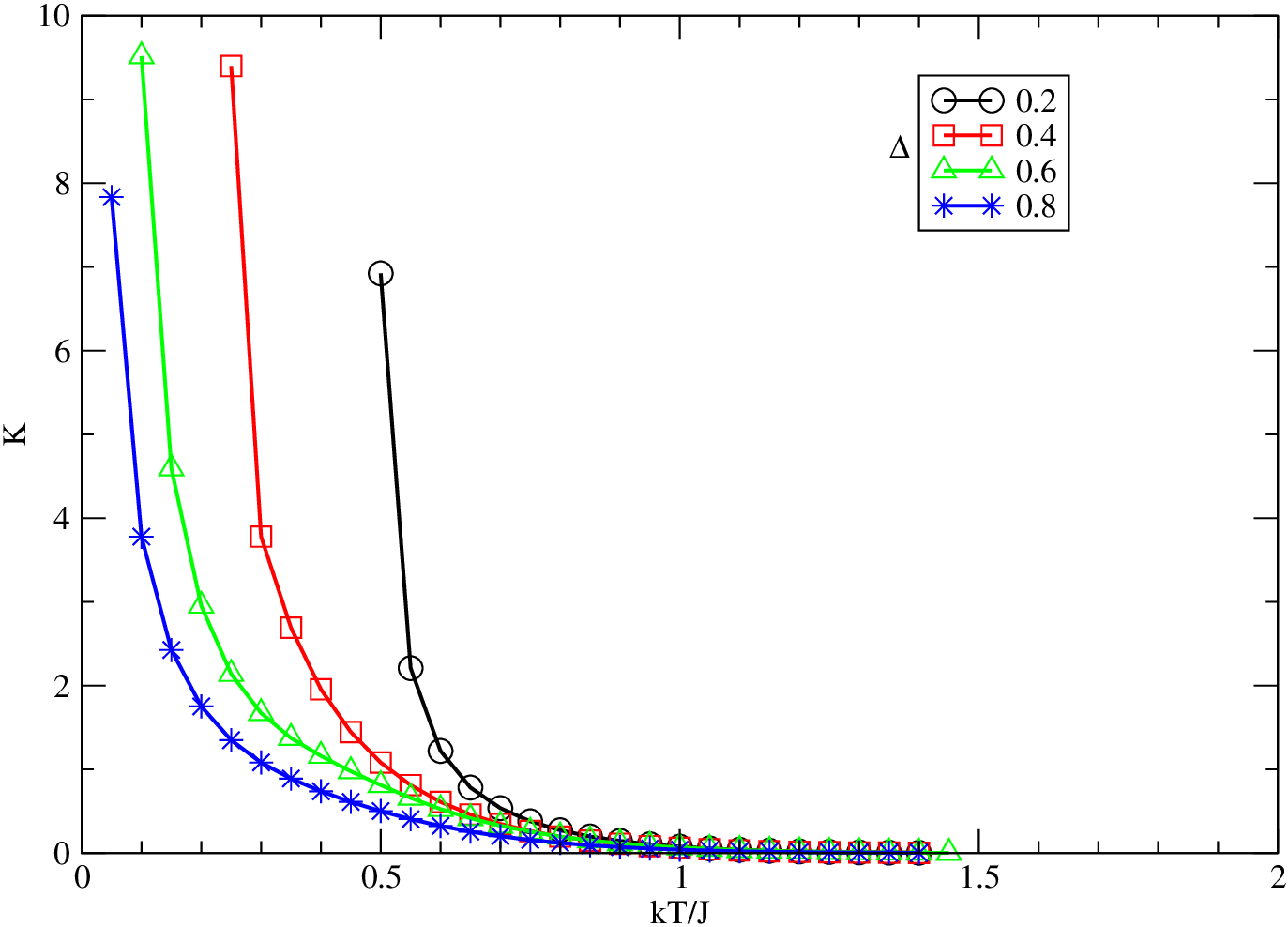}
\end{center}
%\hspace{2pc}%
%\begin{minip-age}[b]{5.5cm}
\caption{\label{fig4} Thermal conductivity versus temperature ($kT/J$) for different anisotropies
in the rung coupling ($\Delta$) and $J_{\bot}/J=2.5$ and $\delta=1.0$. 
A strong dependence on $\Delta$ is clearly observed. }
%\end{minipage}
\end{figure}

We have also studied the effect of anisotropy on the thermal conduction of a spin ladder.
In Fig.\ref{fig3} we plot $K$ versus normalized temperature for different 
values of anisotropies on the leg Hamiltonian, namely $\delta=0.2, 0.4, 0.6, 0.8$
for $J_{\bot}/J=0.2, 1.0, 2.5$. This plot indicates a weak dependence on $\delta$
at very low temperatures. It can be understood from the fact that the singlet-triplet gap is 
practically independent of the leg anisotropy ($\delta$) and thus the triplon density and the thermal transport 
are unaffected. This behavior 
is almost the same for weak coupling $J_{\bot}/J=0.2$, the normal ladder $J_{\bot}/J=1.0$
and the strong coupling limit $J_{\bot}/J=2.5$ as shown in Fig.\ref{fig3}.
All plots presented in Fig.\ref{fig3} show no dependence
on the leg anisotropy. Specially, at the strong coupling limit $J_{\bot}/J >> 1$ the
thermal transport is clearly independent of $\delta$ where all plots fall on each other 
on the whole range of temperature. However, the situation is different for the rung
anisotropy ($\Delta$). Here the rung anisotropy has a direct influence on the singlet-triplet
energy gap and hence on the triplon density as shown in Fig.\ref{fig4} where $K$ is plotted versus 
normalized temperature for different values of the rung anisotropy. The increase of 
$\Delta$ raises the triplon gap which gives lower conductivity at a given temperature. 
Moreover, lower values of $\Delta$ mean weaker interactions on the rungs which consequently 
improve the thermal conductivity up to the limit of the ballistic regime of the
integrable chain. 
In addition, at fixed values of coupling constants which
means fixed triplon density, lower temperature causes less scattering between
triplons and consequently higher values in thermal conductivity. Thus, as 
$T \rightarrow 0$ the scattering of triplons goes to zero which leads
to the divergence of the thermal conductivity.

\section{Summary}
In conclusion, we have presented the temperature dependence of the thermal conductivity 
of an anisotropic spin ladder model due to spin excitation modes. Using a singlet-triplet presentation  
and a Green's function approach the excitation spectrum of the spin ladder has been studied. 
In particular, the effect of anisotropies along the ladder and rung directions have been investigated. 
We have found that the anisotropy along the ladder direction has a major effect on the thermal conductivity, 
while the anisotropy along the leg direction has a minor one.
Also the results show that a decrease of the coupling exchange constant along the rungs 
results to a divergent behavior of the thermal conductivity versus temperature, reminiscent of the 
purely ballistic thermal transport in the Heisenberg spin - 1/2 chain. 

\begin{acknowledgments}
This work was supported in part by the Office of Vice-President for Research
of Sharif University of Technology. 
A.L. acknowledges partial support from the Alexander von Humboldt Foundation
and Max-Planck-Institut f\"ur Physik komplexer Systeme (Dresden-Germany).
P.H.M.v.L. and X.Z. acknowledge the support by the European Commission through the LOTHERM (FP7-238475) project.
The work has been co-financed by the EU (ESF) and Greek national funds through 
the  Operational Program ``Education and Lifelong Learning'' of the NSRF under  
``Funding of proposals that have received a positive evaluation in the 3rd and 4th Call of ERC Grant Schemes''.
\end{acknowledgments}

%\end{minipage}

%\begin{figure}[ht!]
%\begin{center}
%\includegraphics[width=10cm]{11.eps}
%\end{center}
%\hspace{2pc}%
%\begin{minip-age}[b]{5.5cm}
%\caption{\label{fig4} The variation of thermal conductivity along leg direction versus $KT/J$ for 
%different intersite anisotropy in $J_{\bot}/J=0.2$ and $\Delta=0.2$. A monotonic decreasing behavior 
%is readily seen for all cases.}
%\end{minipage}
%\end{figure}

%%%%%%%%%%%%%%%%%%%%%%%%%%%%%%%%%%%%%%%%%%%%%%%%%%%%%%%%

%%%%%%%%%%%%%%%%%%%%%%%%%%%%%%%%%%%%%%%%%%%%%%%%%%%%%%%%%%%%%
\appendix
\section{The explicit expression of $\mathcal{D}(k,k',\epsilon)$}
\setcounter{section}{1}
 In this Appendix, we present the full expression of $\mathcal{D}(k,k',\epsilon)$ which 
has been mentioned in Eq.(\ref{e238}). Let us first define the following relations
\begin{eqnarray}
\phi_{1}&=&l_{2}l_{3}\Big(n_{B}(-l_{2}\omega_{x}(k))-n_{B}(l_{3}\omega_{z}(k+k'))\Big),\nonumber\\
\phi_{2}&=&l_{2}\Big(n_{B}(\omega_{x}(q))-n_{B}(\omega_{x}(k+k'+q))\Big)
\Big(n_{B}(-l_{2}\omega_{x}(k))-n_{B}(-\omega_{x}(k+k'+q)+\omega_{x}(q))\Big),\nonumber\\
\phi_{3}&=&l_{2}\Big(n_{B}(\omega_{z}(q))-n_{B}(\omega_{x}(k+q))\Big)
\Big(n_{B}(\omega_{x}(k+q)-\omega_{z}(q))-n_{B}(l_{2}\omega_{z}(k+k'))\Big),\nonumber\\
\phi_{4}&=&l_{2}\Big(n_{B}(\omega_{x}(q))-n_{B}(\omega_{z}(k+q))\Big)
\Big(n_{B}(\omega_{z}(k+q)-\omega_{x}(q))-n_{B}(l_{2}\omega_{z}(k+k'))\Big),\nonumber\\
\phi_{5}&=&\Big(n_{B}(\omega_{x}(q))-n_{B}(\omega_{x}(k+k'+q))\Big)
\Big(n_{B}(\omega_{z}(q))-n_{B}(\omega_{x}(k+q))\Big)\nonumber\\
&\times&\Big(n_{B}(\omega_{x}(k+q')-\omega_{z}(q))-n_{B}(\omega_{x}(q')-\omega_{x}(k+k'+q))\Big)\nonumber\\
\phi_{6}&=&\Big(n_{B}(\omega_{x}(q))-n_{B}(\omega_{x}(k+k'+q))\Big)
\Big(n_{B}(\omega_{x}(q))-n_{B}(\omega_{z}(k+q))\Big),\nonumber\\
&\times&\Big(n_{B}(\omega_{z}(k+q')-\omega_{x}(q))-n_{B}(\omega_{x}(q')-\omega_{x}(k+k'+q))\Big),\nonumber\\
\lambda_{1}&=&n_{B}(\omega_{z}(q_1))-n_{B}(\omega_{x}(k'+q_1)),\nonumber\\
\lambda_{2}&=&n_{B}(\omega_{x}(q_1))-n_{B}(\omega_{z}(k'+q_1)).\nonumber\\
\label{e239}
\end{eqnarray}
Based on the above definitions, $\mathcal{D}(k,k',\epsilon)$ is given by
\begin{equation}
\mathcal{D}(k,k',\epsilon)=\mathcal{D}_{1}(k,k',\epsilon)+\mathcal{D}_{2}(k,k',\epsilon)+\mathcal{D}_{3}(k,k',\epsilon)+\mathcal{D}_{4}(k,k',\epsilon)+\mathcal{D}_{5}(k,k',\epsilon)+\mathcal{D}_{6}(k,k',\epsilon),
\end{equation}
where
\begin{eqnarray}
\mathcal{D}_{1}(k,k',\epsilon)&=&-4\Big(\sum_{l_{1},l_{2},l_{3}=\pm}l_{1}\phi_{1}Im(1/(\epsilon-l_{1}\omega_{x}(k')+i0^{+}))Im(1/(-\epsilon-l_{2}\omega_{x}(k)-l_{3}\omega_{z}(k+k')-i0^{+}))\nonumber\\&+&\frac{1}{N}\sum_{l_{1}=\pm,l_{2}=\pm,q_1}\lambda_1\phi_{1}Im(1/(\epsilon-\omega_{z}(q_1)+\omega_{x}(k+k'+q_1)+i0^{+}))Im(1/(-\epsilon-l_{1}\omega_{x}(k)-l_{2}\omega_{z}(k+k')-i0^{+}))\nonumber\\&+&\frac{1}{N}\sum_{l_{1}=\pm,l_{2}=\pm,q_1}\lambda_2\phi_{1}Im(1/(\epsilon-\omega_{x}(q_1)+\omega_{z}(k+k'+q_1)+i0^{+}))Im(1/(-\epsilon-l_{1}\omega_{x}(k)-l_{2}\omega_{z}(k+k')-i0^{+}))\Big),\nonumber\\
\mathcal{D}_{2}(k,k',\epsilon)&=&-8\frac{1}{N}\Big(\sum_{l_{1},l_{2}=\pm,q}l_{1}\phi_{2}Im(1/(\epsilon-l_{1}\omega_{x}(k')+i0^{+}))Im(1/(-\epsilon-l_{2}\omega_{x}(k)-\omega_{x}(q)+\omega_{x}(k+k'+q)-i0^{+}))\nonumber\\&+&\frac{1}{N}\sum_{l_{1}=\pm,q,q_1}\lambda_1\phi_{2}Im(1/(\epsilon-\omega_{z}(q_1)+\omega_{x}(k'+q_1)+i0^{+}))Im(1/(-\epsilon-l_{1}\omega_{x}(k)-\omega_{x}(q)+\omega_{x}(k+k'+q)-i0^{+}))\nonumber\\&+&\frac{1}{N}\sum_{l_{1}=\pm,q,q_1}\lambda_2\phi_{2}Im(1/(\epsilon-\omega_{x}(q_1)+\omega_{z}(k'+q_1)+i0^{+}))\nonumber\\&\times&Im(1/(-\epsilon-l_{1}\omega_{x}(k)-\omega_{x}(q)+\omega_{x}(k+k'+q)-i0^{+}))\Big),\nonumber\\
\mathcal{D}_{3}(k,k',\epsilon)&=&-4\frac{1}{N}\Big(\sum_{l_{1},l_{2}=\pm,q}l_{1}\phi_{3}Im(1/(\epsilon-l_{1}\omega_{x}(k')+i0^{+}))Im(1/(-\epsilon-l_{2}\omega_{z}(k+k')-\omega_{z}(q)+\omega_{x}(k+q)-i0^{+}))\nonumber\\&+&\frac{1}{N}\sum_{l_{1}=\pm,q,q_1}\lambda_1\phi_{3}Im(1/(\epsilon-\omega_{z}(q_1)+\omega_{x}(k'+q_1)+i0^{+}))Im(1/(-\epsilon-l_{2}\omega_{z}(k+k')-\omega_{z}(q)+\omega_{x}(k+q)-i0^{+}))\nonumber\\&+&\frac{1}{N}\sum_{l_{1}=\pm,q,q_1}\lambda_2\phi_{3}Im(1/(\epsilon-\omega_{x}(q_1)+\omega_{z}(k'+q_1)+i0^{+}))\nonumber\\&\times&Im(1/(-\epsilon-l_{2}\omega_{z}(k+k')-\omega_{x}(q)+\omega_{x}(k+q)-i0^{+}))\Big),\nonumber\\
\mathcal{D}_{4}(k,k',\epsilon)&=&-4\frac{1}{N}\Big(\sum_{l_{1},l_{2}=\pm,q}l_{1}\phi_{4}Im(1/(\epsilon-l_{1}\omega_{x}(k')+i0^{+}))Im(1/(-\epsilon-l_{2}\omega_{x}(k+k')-\omega_{z}(q)+\omega_{z}(k+q)-i0^{+}))\nonumber\\&+&\frac{1}{N}\sum_{l_{1}=\pm,q,q_1}\lambda_1\phi_{4}Im(1/(\epsilon-\omega_{z}(q_1)+\omega_{x}(k'+q_1)+i0^{+}))Im(1/(-\epsilon-l_{2}\omega_{x}(k+k')-\omega_{z}(q)+\omega_{z}(k+q)-i0^{+}))\nonumber\\&+&\frac{1}{N}\sum_{l_{1}=\pm,q,q_1}\lambda_2\phi_{4}Im(1/(\epsilon-\omega_{x}(q_1)+\omega_{z}(k'+q_1)+i0^{+}))\nonumber\\&\times&Im(1/(-\epsilon-l_{2}\omega_{z}(k+k')-\omega_{x}(q)+\omega_{x}(k+q)-i0^{+}))\Big),\nonumber\\
\mathcal{D}_{5}(k,k',\epsilon)&=&-4\frac{1}{N}\Big(\sum_{l_{1}=\pm,q,q_1}l_{1}\phi_{5}\nonumber\\&\times&Im(1/(\epsilon-l_{1}\omega_{x}(k')+i0^{+}))Im(1/(-\epsilon+\omega_{x}(k+k'+q)-\omega_{z}(q')-\omega_{z}(q)+\omega_{x}(k+q')-i0^{+}))\nonumber\\&+&\frac{1}{N}\sum_{l_{1}=\pm,q,q_1,q'}\lambda_1\phi_{5}Im(1/(\epsilon-\omega_{z}(q_1)+\omega_{x}(k'+q_1)+i0^{+}))\nonumber\\&\times&Im(1/(-\epsilon+\omega_{x}(k+k'+q)-\omega_{z}(q')-\omega_{z}(q)+\omega_{x}(k+q')-i0^{+}))\nonumber\\&+&\frac{1}{N}\sum_{l_{1}=\pm,q,q_1,q'}\lambda_2\phi_{5}Im(1/(\epsilon-\omega_{x}(q_1)+\omega_{z}(k'+q_1)+i0^{+}))\nonumber\\&\times&Im(1/(-\epsilon+\omega_{x}(k+k'+q)-\omega_{z}(q')-\omega_{z}(q)+\omega_{x}(k+q')-i0^{+}))\Big),\nonumber\\
\mathcal{D}_{6}(k,k',\epsilon)&=&-4\frac{1}{N}\Big(\sum_{l_{1}=\pm,q,q_1}l_{1}\phi_{6}Im(1/(\epsilon-l_{1}\omega_{x}(k')+i0^{+}))\nonumber\\&\times&Im(1/(-\epsilon+\omega_{x}(k+k'+q)-\omega_{x}(q')-\omega_{z}(q)+\omega_{z}(k+q')-i0^{+}))\nonumber\\&+&\frac{1}{N}\sum_{l_{1}=\pm,q,q_1}\lambda_1\phi_{6}Im(1/(\epsilon-\omega_{z}(q_1)+\omega_{x}(k'+q_1)+i0^{+}))\nonumber\\&\times&Im(1/(-\epsilon+\omega_{x}(k+k'+q)-\omega_{x}(q')-\omega_{z}(q)+\omega_{z}(k+q')-i0^{+}))\nonumber\\&+&\frac{1}{N}\sum_{l_{1}=\pm,q,q_1,q'}\lambda_2\phi_{6}Im(1/(\epsilon-\omega_{x}(q_1)+\omega_{z}(k'+q_1)+i0^{+}))\nonumber\\&\times&Im(1/(-\epsilon+\omega_{x}(k+k'+q)-\omega_{x}(q')-\omega_{z}(q)+\omega_{z}(k+q')-i0^{+}))\Big).
\end{eqnarray}

\end{document}